# Total angular momentum sorting in the telecom infrared with silicon Pancharatnam-Berry transformation optics

G. RUFFATO,[1,2,†] P. CAPALDO,[2,3] M. MASSARI,[2,3] E. MAFAKHERI,[1,2] AND F. ROMANATO[1,2,3,*]

[1]*Department of Physics and Astronomy 'G. Galilei', University of Padova, via Marzolo 8, 35131 Padova, Italy*
[2]*LANN, Laboratory for Nanofabrication of Nanodevices, EcamRicert, Corso Stati Uniti 4, 35127 Padova, Italy*
[3]*CNR-INFM TASC IOM National Laboratory, S.S. 14 Km 163.5, 34149 Basovizza, Trieste, Italy*
[†]*gianluca.ruffato@unipd.it*
[*]*filippo.romanato@unipd.it*

**Abstract:** Parallel sorting of orbital angular momentum (OAM) and polarization has recently acquired paramount importance and interest in a wide range of fields ranging from telecommunications to high-dimensional quantum cryptography. Due to their inherently polarization-sensitive optical response, optical elements acting on the geometric phase prove to be useful for processing structured light beams with orthogonal polarization states by means of a single optical platform. In this work, we present the design, fabrication and test of a Pancharatnam-Berry optical element in silicon implementing a *log-pol* optical transformation at 1310 nm for the realization of an OAM sorter based on the conformal mapping between angular and linear momentum states. The metasurface is realized in the form of continuously-variant subwavelength gratings, providing high-resolution in the definition of the phase pattern. A hybrid device is fabricated assembling the metasurface for the geometric phase control with multi-level diffractive optics for the polarization-independent manipulation of the dynamic phase. The optical characterization confirms the capability to sort orbital angular momentum and circular polarization at the same time.

## 1. Introduction

During the last decades, different solutions have been introduced and engineered in order to carry many signals through the same optical link. Basically, they relied on tailoring the different degrees of freedom of light in order to encode more information on the same carrier: time, amplitude/phase, wavelength, polarization [1]. However, also the combination of these techniques is limited, and it is not sufficient to enhance the transmission capacity beyond the physical limit of single-mode fibers (Shannon limit) [2]. More recently, the manipulation of the phase and intensity spatial distribution of light beams has become paramount as a complementary and promising solution to satisfy the restless request for traffic capacity. In particular, mode-division multiplexing (MDM) suggests the exploitation of the several orthogonal distributions that light can assume in a multi-mode medium in order to increase proportionally the number of available channels at the same wavelength. Light beams carrying orbital angular momentum (OAM), endowed with helical phase-fronts and annular intensity patterns [3], have been demonstrated to provide a robust and viable solutions for applications either in free-space or optical fibers [4, 5], and have gained increasing attention both in academy and industry.

The pivotal components of an optical link based on OAM-MDM is represented by the multiplexer and the demultiplexer, i.e. the devices used to put together and sort, respectively, optical beams with different values of OAM. The method based on the conformal mapping between linear and orbital angular momentum states has known increasing attention and application due to its efficiency and scalability. The mapping is executed by two optical elements, in sequence: the first performing a *log-pol* coordinate transformation (the unwrapper), and the second correcting the introduced phase distortion (the phase-corrector). After its first demonstration with spatial light modulators (SLMs) [6], the sorter was subsequently replaced by refractive optical components [7] for efficiency reasons and upgraded in terms of resolution [8, 9]. In the quest for miniaturization, a diffractive version [10] was proposed, exhibiting an unprecedented combination of compactness [11] and optical performance [12]. However, the phase added by those phase elements is essentially based on a transverse modulation of the optical path-length experienced by the cross-section of the beam. Other types of phase-modulating mechanisms could also be employed for this purpose such as those that transfer a geometric phase [13]. This is achievable by moving from a patterned isotropic substrate to a spatially-variant anisotropic medium, as in the so-called Pancharatnam-Berry optical elements (PBOE), which are able to transfer the desired phase pattern by locally manipulating the polarization state of the impinging light. As a matter of fact, by properly structuring the medium to induce an effective artificial birefringence [14], a phase delay equal to twice the rotation angle of the local fast axis is imparted point by point. Since the sign of the experienced phase delay depends on the input circularly-polarized state, the optical response is inherently polarization-sensitive and many PBOEs have been developed purposely in order to discriminate the optical operations of the optics on the basis of the input polarization state [15-21].

Recently, a theoretical proposal introduced a Pancharatnam-Berry OAM sorter based on transformation optics exploiting this relation with the input polarization as a means of additionally sorting optical spin states [22]. Lately, the implementation of such a total optical angular momentum sorter was reported, with custom-made liquid-crystal devices as the required PBOEs [23-25], exploiting the birefringence properties of the photo-aligned polymers. However, those demonstrations were limited to the visible range, and the fabrication technique was limited in resolution and not easily compatible with mass-production techniques.

Here we present, for the first time, the fabrication of a metasurface performing a *log-pol* optical transformation for OAM-MDM at the telecom infrared at 1310 nm. The choice of operating in the infrared opens to the usage of silicon as substrate material, which has attracted an increasing interest due to its large dielectric constant, low absorption, and maturity of nanostructuring techniques. Among the many different combinations of optical design and materials, subwavelength silicon metasurfaces are reserving a particular place for the design and fabrication of infrared optical elements which are able to add tailored yet opposite phases to orthogonal circularly-polarized beams accompanied by a polarization conversion [26-28]. In recent papers [29, 30], we have demonstrated the possibility to implement a discretized phase pattern in the form of a meta-pixel matrix of rotated sub-wavelength gratings. However, in the case of phase maps where high gradients and singularities are present, a discretization of the phase pattern over a coarse mesh could dramatically affect the optical performance and the diffraction efficiency. One possible solution consists in designing continuously space-variant metasurfaces, by determining the local direction and period of a sub-wavelength grating using vectorial optics in order to obtain any desired continuous polarization change, hence completely suppressing any diffraction effect arising from polarization discontinuity. The method presented by Hasman's group [16, 17] was here extended to the design of a continuously-variant subwavelength-grating unwrapper in silicon.

We optically characterized the fabricated sorter by feeding it with beams carrying well-defined values of OAM and circular polarizations, demonstrating the expected capability to sort spin and orbital angular momentum with the same optical platform. The use of silicon will enable the integration of metasurface devices into existing technology platforms, due to the mainstream industrialization and mature development of silicon in electronics, on-chip photonics, and microelectromechanical systems.

## 2. Theory and design

### 2.1 Transformation optics with Pancharatnam-Berry optical elements

OAM sorters based on transformation optics demonstrated how OAM states, associated to azimuthal phase gradients, can be converted into transverse momentum states, associated to linear phase gradients, through a *log-pol* optical transformation $(x, y) \rightarrow (u, v)$, followed by a phase correction. An azimuthal phase gradient $\exp(i\ell\varphi)$ can be transformed into a linear phase gradient $\exp(iKv)$, by applying a coordinate change $v=a\cdot\arctan(y/x)$, being $K=2\pi\ell/L$, $L=2\pi a$. The conformal conditions impose the relation $u=-a\cdot\ln(r/b)$, defining the position of the unwrapped field in the $u$ direction. The optical element performing this transformation, the so-called un-wrapper, is endowed with a phase function given by [6]:

$$\Omega_{UW}(x, y) = \frac{2\pi a}{\lambda f_1}\left[y\cdot\arctan\left(\frac{y}{x}\right) - x\cdot\ln\left(\frac{\sqrt{x^2+y^2}}{b}\right) + x\right] - \frac{2\pi}{\lambda}\frac{x^2+y^2}{2f_1} \quad (1)$$

where the parameter $a$ is chosen in order to ensure that the azimuthal angle range $(0, 2\pi)$ is mapped onto the full width $L=2\pi a$, while the parameter $b$ is optimized for the size of the sorter and can be chosen independently to control the transformed-beam position in the orthogonal direction. In order to compensate phase distortions due to propagation and restore the linear phase gradient, a second optical element is needed, placed at a distance equal to $f_1$, whose phase function is given by:

$$\Omega_{PC}(u, v) = -\frac{2\pi ab}{\lambda f_1}\exp\left(-\frac{u}{a}\right)\cos\left(\frac{v}{a}\right) - \frac{2\pi}{\lambda}\frac{u^2+v^2}{2f_1} \quad (2)$$

where a Fresnel correction has been included. If a lens with focal length $f_2$ is placed after the phase-corrector in *f-f* configuration, the transformed beam is focused onto a specified lateral position, proportionally to the OAM content $\ell$ according to:

$$\Delta s = \frac{\lambda f_2}{2\pi a}\ell \quad (3)$$

Now we consider a Pancharatnam-Berry (PB) version of the un-wrapper, assuming the phase function in Eq. (1) valid for illumination under right-handed circular polarization. If the same element is illuminated under left-handed circular polarization, the beam experiences the opposite phase:

$$\Omega_{UW}^{(-)}(x, y) = -\frac{2\pi a}{\lambda f_1}\left[y\cdot\arctan\left(\frac{y}{x}\right) - x\cdot\ln\left(\frac{\sqrt{x^2+y^2}}{b}\right) + x\right] + \frac{2\pi}{\lambda}\frac{x^2+y^2}{2f_1} \quad (4)$$

Due to the change in sign in the quadratic term, it is worth noting that the beam experiences defocusing. Then, the quadratic term cannot be included in the PB version of the optical element, and the focusing operation should be implemented with a polarization-insensitive optical element operating on the dynamic phase, i.e. a bulk lens or a diffractive counterpart. The sign inversion in the un-wrapper phase function corresponds to a reversed unwrapping process $v'=-a\cdot\arctan(y/x)=-v$, in conjunction with a change in sign $u'=a\cdot\ln(r/b)=-u$, imposed by the conformal condition [22]. Therefore, the corresponding phase-corrector is given by the transformation $(u,v) \rightarrow (-u,-v)$. As expressed by the change in sign of the coordinate *u*, the two orthogonal polarizations are focused at opposite, i.e. distinct, positions on the phase-corrector. This can be understood also considering the linear term in the un-wrapper, which make the two circular polarizations experience opposite tilts and therefore be spatially separated on the Fourier plane. Then the phase-corrector operates separately on each one, and we can properly account for the inherent difference in the way the elements act on the opposite circular polarizations. As a consequence, the phase-corrector exhibits a discontinuity at *u*=0, and its phase function is given by:

$$\Omega_{PC}(u,v) = -\frac{2\pi ab}{\lambda f}\exp\left(-\frac{|u|}{a}\right)\cos\left(\frac{v}{a}\right) + \alpha u \tag{5}$$

where an additional grating term with spatial frequency *α* has been added to provide separation between the two polarizations in far-field upon propagation.

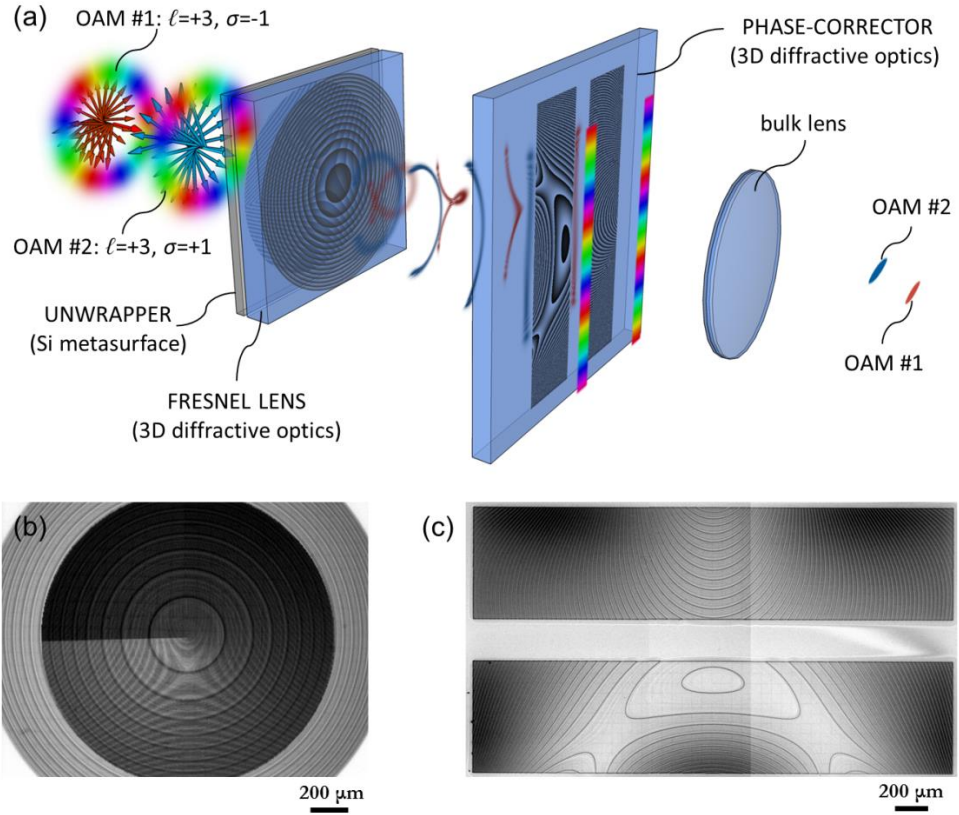

Fig. 1. (a) Scheme of the fabricated sorter performing parallel demultiplexing of polarization and orbital angular momentum with a Pancharatnam-Berry transformation optics. The input circularly-polarized OAM states illuminate the un-wrapper metasurface performing *log-pol* optical transformation. A Fresnel lens is coupled with the first element in order to focus the unwrapping beam on the second element independently of its polarization state. The two polarizations illuminate two distinct zones of the second element and can be processed separately with a second multi-level diffractive optics performing phase correction. Finally, a lens is used in *f-f* configuration to complete the sorting process. (b) Microscope image of the hybrid focusing un-wrapper formed by coupling a multi-level Fresnel lens (on the top) with a metasurface un-wrapper (SEM inspection in Fig. 5). (c) Inspection of the multi-level phase-corrector.

The separation between the orthogonally-polarized components enabled by the first PBOE thereby allows for a device capable of sorting optical angular momentum eigenstates and, in turn, of handling an increased amount of information channels that can be used in practical applications. It is worth noting that the polarization sorting occurs in the unwrapping stage, and the two polarizations illuminate the phase-corrector in two distinct zones. Therefore, in principle, there is no strict need to realize the phase-corrector in a metasurface form as well, and it can be fabricated in the form of a multi-level diffractive optics [10].

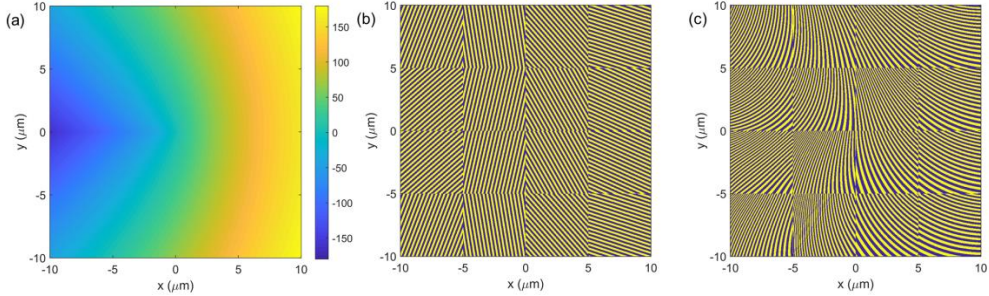

Fig. 2. (a) Un-wrapper phase map defined by Eq. (1) with parameters $a=400$ μm, $b=150$ μm, $f_1=2.54$ cm, without lens term. (b) Implementation in the form of sub-wavelength grating Pancharatnam-Berry optical element with phase discretization over a pixel mesh with size 5 μm. (c) Implementation in the form of continuously-variant subwavelength gratings over the same mesh. It is worth noting the higher resolution in the geometric phase definition, given by the local grating vector rotation, and its continuity at the pixel boundaries.

## 2.2 Design of continuously-variant subwavelength-grating un-wrapper

The local anisotropy of the metasurface un-wrapper was induced by properly engineering a sub-wavelength grating pattern. As described in [29], the effective refractive index is lower along the grating vector direction, and the grating vector orientation is exploited to control the transferred geometric phase. In order to define the phase pattern with the highest resolution, the grating vector orientation should vary continuously, defining a spatially-variant grating structure (see Fig. (2)). To ensure the continuity of the grating pattern of the PB un-wrapper, we impose the following condition on the grating vector:

$$\nabla \times \vec{G} = 0 \qquad (6)$$

where the grating vector is spatially variant and defined as:

$$\vec{G}(x,y) = G(x,y)\begin{bmatrix} \cos\vartheta(x,y) & \sin\vartheta(x,y) \end{bmatrix} \qquad (7)$$

$\vartheta(x,y)$ being the grating rotation angle, equal to half the geometric phase. After substituting the last expression into Eq. (6) we get a system of partial differential equations. We obtain the following conditions on the grating-vector partial derivatives:

$$\begin{cases} \dfrac{\partial G(x,y)}{\partial x} = -G(x,y)\dfrac{\partial \vartheta(x,y)}{\partial y} \\ \dfrac{\partial G(x,y)}{\partial y} = G(x,y)\dfrac{\partial \vartheta(x,y)}{\partial x} \end{cases} \qquad (8)$$

By substituting the spatial dependency $\vartheta(x,y)$ given by the desired optical element, and then integrating, a spatial dependency $G(x,y)$ of the grating-vector modulus is obtained. Once the grating vector $G(x,y)$ has been determined, the grating structure can be outlined by introducing the grating function $\gamma(x,y)$, whose existence is assured by the conservative condition in Eq. (6), calculated by integration over an arbitrary path:

$$\nabla \gamma = \vec{G} \qquad (9)$$

Finally, the binarization of the grating function is performed by imposing the following period-dependent Lee-type condition:

$$T(x,y) = \Theta\{\cos\gamma(x,y) - \cos(\pi q(\Lambda(x,y)))\} \quad (10)$$

where $\Theta$ is the Heaviside function, $q(\Lambda(x,y))$ is the duty-cycle calculated for the grating wavelength at the position $(x,y)$: $\Lambda(x,y)=2\pi/G(x,y)$. By applying the previous method on the un-wrapper phase pattern given by Eq. (1) without the focusing term:

$$\vartheta_{UW}(x,y) = \frac{\pi a}{\lambda f_1}\left[y\cdot\arctan\left(\frac{y}{x}\right) - x\cdot\ln\left(\frac{\sqrt{x^2+y^2}}{b}\right) + x\right] \quad (11)$$

the following analytical function is obtained for the un-wrapper grating vector $G_{UW}(x,y)$:

$$G_{UW}(x,y) = G_0 \cdot \exp\left\{\frac{\pi a}{\lambda f_1}\left[x\cdot\arctan\left(\frac{x}{y}\right) - y\cdot\ln\left(\frac{\sqrt{x^2+y^2}}{b}\right) + y - \frac{\pi}{2}x\right]\right\} \quad (12)$$

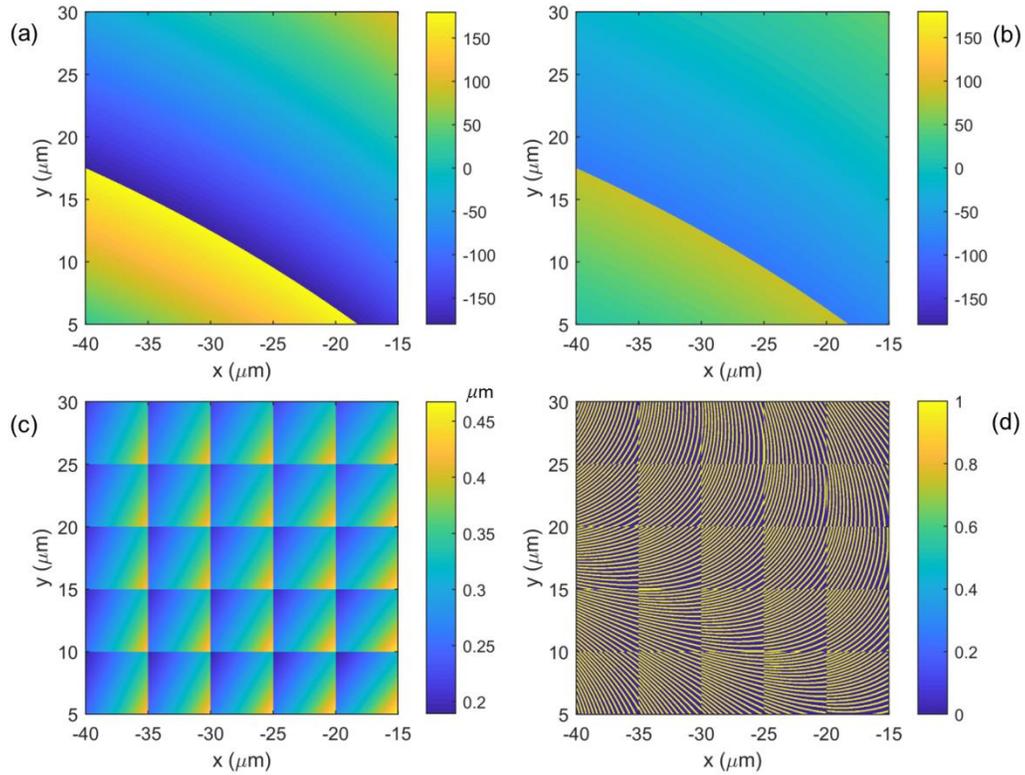

Fig. 3. Conversion of a phase map (a) into a spatially-variant subwavelength digital grating (d). (a) Phase map of a selected zone of the un-wrapper phase pattern. (b) Rotation angle of the grating fast axis with respect to the *x*-axis. The rotation angle is half the phase function in (a). (c) Period of the subwavelength grating over the selected zone. The zone is discretized into meta-pixels of lateral size 5 μm. Inside each meta-pixel, the grating period can vary between an upper limit, close to the structural cut-off, and a lower limit imposed by the fabrication process. (d) After imposing a period-dependent Lee-type digitalization (Eq. (10)), the subwavelength binary grating is finally obtained.

However, the grating period is not allowed to vary without limits. As a matter of fact, an upper value ($\Lambda_{co}$) and a lower value ($\Lambda_m$) threshold should be defined, where the upper value is given by the grating structural cut-off [31], in order to preserve the metasurface behavior. For instance, in case of silicon gratings, at a wavelength of 1310 nm and normal incidence, the structural cut-off is around 400 nm. The lower value $\Lambda_m$ is imposed by the lithographic limitations, and it is given by the minimum grating-width obtainable by etching the required thickness (a reasonable value for $\Lambda_m$ can be around 150 nm for a grating depth around 500 nm). The whole phase pattern is divided into a matrix of zones: inside each zone the grating potential is calculated, as described above, allowing the period to vary between $\Lambda_m$ and $\Lambda_{co}$, as shown in Fig. 3. As imposed in Eq. (10), a spatially-varying duty-cycle is necessary in order to preserve a phase delay $\delta=\pi$ between the grating ordinary and extraordinary axes over the whole surface. Numerical simulations have been performed implementing Rigorous Coupled-Wave Analysis (RCWA) [32-34] for a binary silicon grating in air at 1310 nm, in order to extract the optimal configurations of duty-cycle and period providing $\pi$ retardation. As described in [29], two distinct optimal-configuration branches can be obtained.

As a result of the design algorithm, the grating rotation angle, equal to half the transferred geometric phase, is continuous at the boundary between adjacent zones (see Fig. 2 and 3). Inside each zone the grating vector varies continuously with both grating period values inside the imposed range and a local rotation angle which is given by Eq. (11).

As already mentioned, since the polarization sorting occurs in the unwrapping stage and the two polarizations can be processed separately with the phase-corrector, we fabricated it in the form of a multi-level diffractive optical element. In addition, the focusing term in Eq. (1) was implemented in the form of a multi-level Fresnel lens which was coupled with the metasurface un-wrapper directly as outlined in Fig. (1). A Fresnel correction was added to the phase-corrector phase function, as is included in Eq. (2).

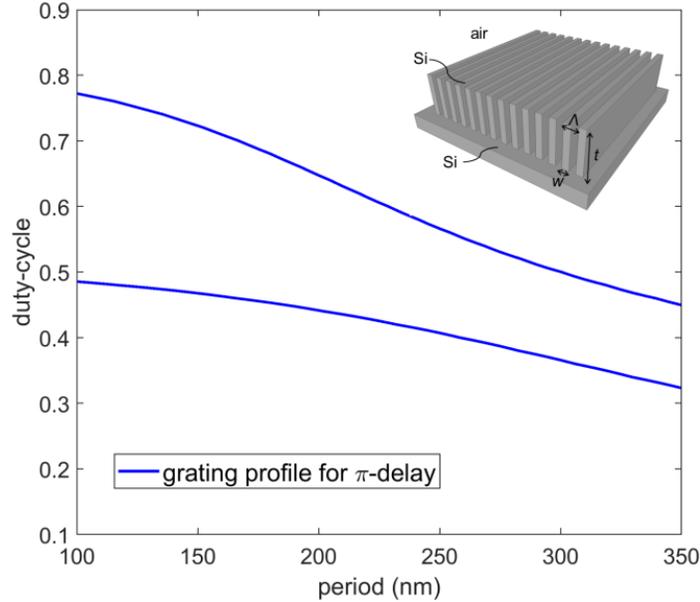

Fig. 4. Configurations of duty cycle and period for a silicon subwavelength grating, providing a $\pi$-delay retardation between TE and TM polarizations, for a wavelength of 1310 nm at normal incidence. Numerical data obtained with rigorous coupled-wave analysis (RCWA) for a grating thickness of 540 nm.

## 3. Fabrication

The sorter proposed in this work represents a hybrid implementation which allows to control at the same time both the geometric and dynamic phase of the input light. The fabrication of the silicon un-wrapper was performed with two distinct processes, in sequence: fabrication of the resist mask with electron-beam lithography (EBL) of a resist film spun over the silicon substrate, and transfer of the mask profile thereon with inductively coupled plasma–reactive ion etching (ICP-RIE). Due to the polarization-sensitive optical response of metasurfaces, the focusing term in Eq. (1) could not be included in the silicon un-wrapper, therefore a multi-level Fresnel lens was fabricated to be coupled directly onto the surface of the first element. The diffractive optical elements were fabricated with electron-beam lithography of a resist layer spun over a glass substrate. The same technique was chosen to fabricate the phase-corrector.

### *3.1 Metasurface fabrication in silicon*

For the fabrication of sub-wavelength grating with high aspect ratio a stamp process has been considered. EBL is the ideal technique to transfer the computational patterns from a digital stored format to an imaging layer with high-resolution profiles [35]. The original EBL pattern was transformed in an imprinting mold for subsequent imprinting replica and ICP-RIE etching to achieve the final sample.

A JBX-6300FS JEOL EBL machine, 12 MHz, 5 nm lithographic resolution, working at 100 kV with a current of 100 pA was used. A 170 nm layer of AR-P6200.09 resist (Allresist GmbH) was spun on a silicon substrate.

The resist pattern was transferred into the silicon substrate by means of STS MESC MULTIPLEX Reactive Ion Etching (RIE) plasma etching in Inductively Coupled Plasma-Reactive Ion Etching (ICP-RIE) configuration working at 13.56 MHz frequency.

The Thermal-NanoImprint Lithography process for the master fabrication was conducted using a Paul-Otto Weber hydraulic press with heating/cooling plates [36, 37]. The pattern generated by EBL is transformed into master stamp [38, 39]. The T-NIL process is carried out at 80°C for an imprinting time of 20 minutes at a pressure of 100 bar, followed by cooling down at 35°C, lower than resist glass transition temperature. For the fabrication of the final samples (Fig. 5), the T-NIL process was conducted at 80°C at a pressure of 100 bar for 10 minutes followed by the same cooling down. Finally, the ICP-RIE etching was performed to remove the residual layer and hence reach the required grating thickness.

In Fig. 5 scanning electron microscopy (SEM) inspection of the fabricated silicon un-wrapper is shown. The phase pattern given by Eq. (1), with parameters $a$=400 μm, $b$=150 μm, $f_1$=2.54 cm, and without lens contribution, has been pixelated into meta-pixel with size 5 μm. Inside each pixel, the phase pattern has been converted into a continuously-variant sub-wavelength grating using the numerical procedure described above.

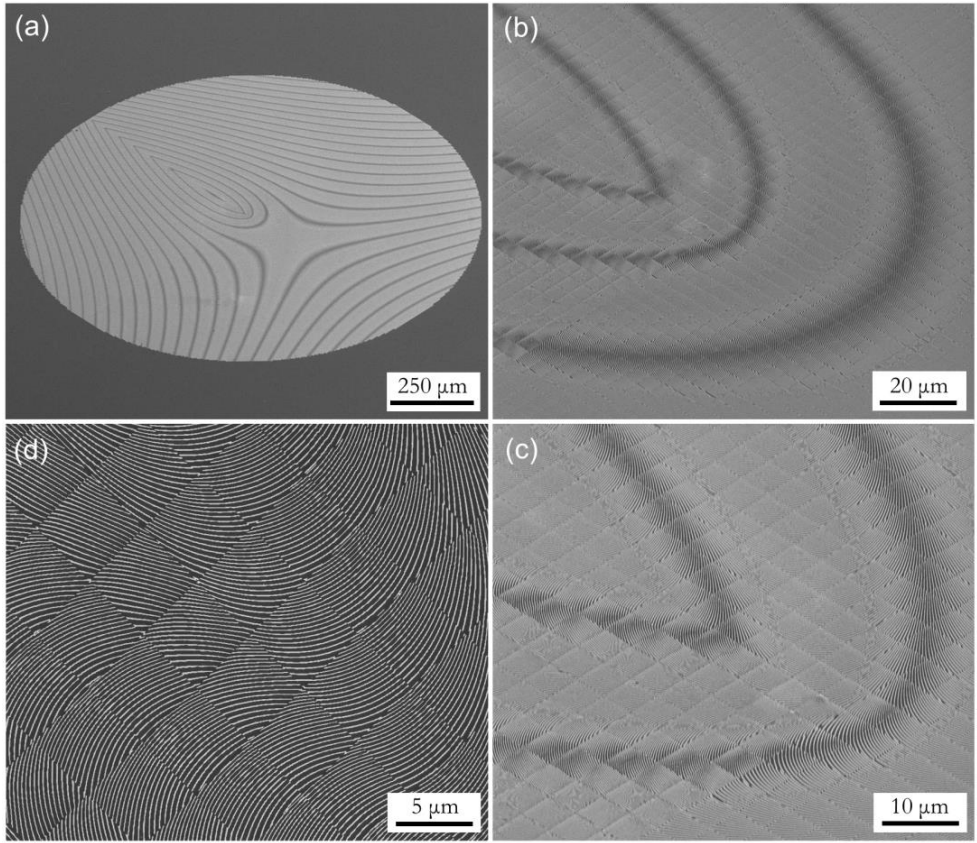

Fig. 5. SEM inspections of the metasurface un-wrapper fabricated in the form of continuously variant sub-wavelength gratings on a silicon substrate. Tilted top view (a) and details at higher magnifications of the central zone (b-c). (d) Top view of a 25 μm x 25 μm zone: it is worth appreciating the continuity of the grating-vector rotation angle at the meta-pixel boundaries. Inside each pixel, the grating vector varies continuously and the duty-cycle changes according to the trend in Fig. (4). Design parameters for the un-wrapper phase function: $a$=400 μm, $b$=150 μm, $f_1$=2.54 cm.

*3.2 3D diffractive optics fabrication*

The Fresnel lens and the phase-corrector have been fabricated as surface-relief phase-only diffractive optics with electron-beam lithography. By locally controlling the released electronic dose, a different dissolution rate is induced in each zone of the exposed polymer, giving rise to different resist thicknesses after development. A dose-depth correlation curve (contrast curve) is required to establish the correct electron-dose to assign to each zone in order to obtain the desired resist thickness. By using custom numerical codes, the transmission phase function of the simulated optics was converted into a 3D multilevel structure, whose local thickness is proportional to the theoretical phase delay, which was in turn transformed into a map of electronic doses.

A dose correction for compensating the proximity effects is applied, in order both to match layout depth with the fabricated relief and to obtain a good shape definition, especially in correspondence of the $2\pi$-phase discontinuities. The optical elements have been fabricated by patterning a 3-μm thick layer of poly(methyl methacrylate) (PMMA) resist, using a JBX-6300FS JEOL EBL machine, 12 MHz, in high-resolution mode, generating at 100 KeV and 100 pA an electron-beam. Few nanometers of gold were sputtered on the surface of the transparent sample to ensure both a better determination of the beam focus and to improve the electron discharge during the exposure. After the exposure, the resist was developed in a temperature-controlled developer bath (deionized water: isopropyl alcohol (IPA) 3:7) for 60 s. After development, the optical elements were gently rinsed in deionized water and blow-dried under nitrogen flux.

At the experimental wavelength of the laser ($\lambda$ = 1310 nm), the resist refractive index was assessed to be $n_R$ = 1.645, as measured by analysis with spectroscopic ellipsometry. For a phase pattern $\Omega(x, y)$, the depth $d(x,y)$ of the resist zone after development, for normal incidence in air, is given by:

$$d(x, y) = \frac{\lambda}{n_R(\lambda) - 1} \frac{2\pi - \Omega(x, y)}{2\pi} \tag{13}$$

The total depth of the relief pattern results 2031 nm, with a thickness step of $\Delta d$ = 8 nm for 256 phase levels (Fig. 6). A SEM inspection of the fabricated sample is shown in Fig. 6. It is worth noting the smoothness of the surface and the steepness in correspondence of $2\pi$ phase jumps.

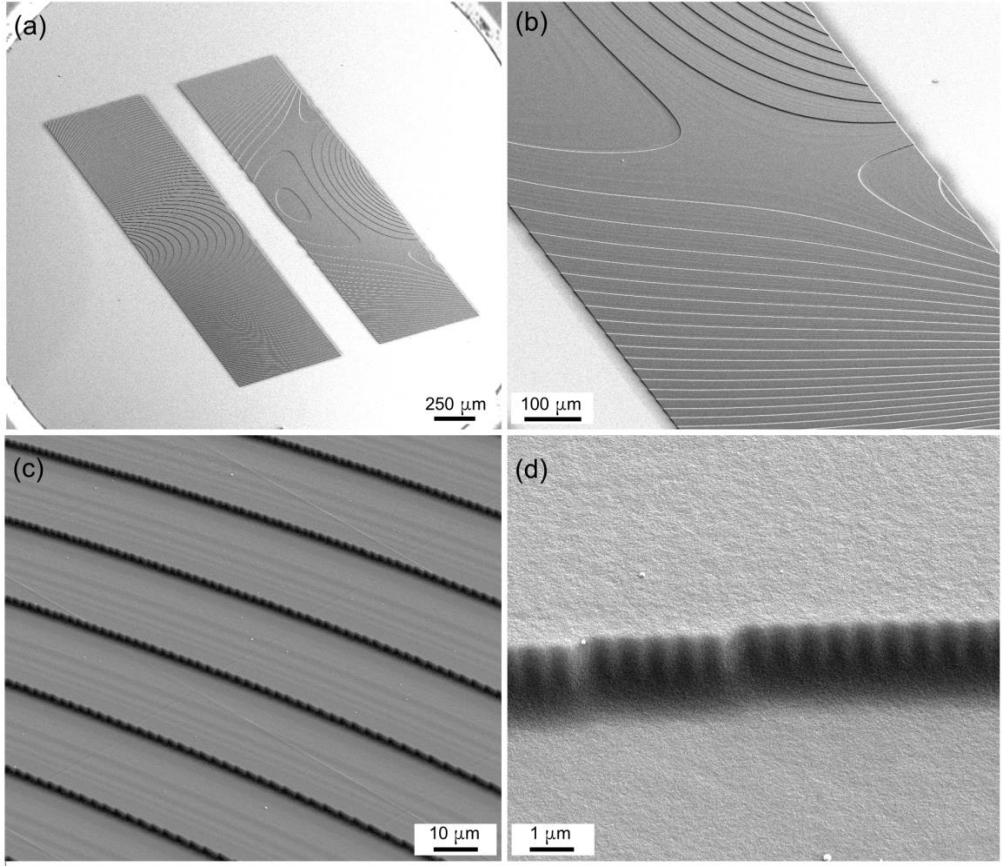

Fig. 6. SEM inspection of the multi-level diffractive phase-corrector and details at higher magnifications. Design parameters: $a$=400 μm, $b$=150 μm, $f_1$=2.54 cm, $\alpha$=0.1 μm$^{-1}$. It is worth noting in (a) the piecewise definition of the phase pattern.

## 4. Optical characterization and results

The optical behavior of the sorting scheme has been tested with the experimental setup depicted in Fig. 7. The input OAM beams were generated with a LCoS spatial light modulator (X13267-08, Hamamatsu, pixel pitch 12.5 μm) using a phase and amplitude modulation technique [40]. The output of a DFB laser (λ=1310 nm) was collimated at the end of the single mode fiber with an aspheric lens with focal length $f_F$=7.5 mm (A375TM-C, Thorlabs), linearly polarized and expanded with a first telescope ($f_1$=3.5 cm, $f_2$=10.0 cm) before illuminating the display of the SLM. Then, a 4-$f$ system ($f_3$=20.0 cm, $f_4$=12.5 cm) with an aperture in the Fourier plane was used to isolate the first-order encoded mode. A 50:50 beam-splitter was used to split the beam and check the input beam profile with a first camera (WiDy SWIR 640U-S, pixel pitch 15 μm). A sequence of linear polarizer (LPIREA100-C, Thorlabs) and quarter-wave plate (WPQ10M-1310, Thorlabs) was used in order to set the desired circular polarization state.

The circularly-polarized OAM beam illuminated the patterned zone of the silicon sample, mounted on a 6-axis kinematic mount (K6XS, Thorlabs). On the surface of the silicon substrate, on the etched side, a Fresnel lens was centered and attached. Then the transformed unwrapping beam was focused, independently of the input polarization, on the phase-corrector, mounted and aligned onto a second 6-axis mount (K6XS, Thorlabs), whose position along the optical axis could be adjusted using a micrometric translator (TADC-651, Optosigma). The far-field was collected by a second camera (WiDy SWIR 640U-S) placed at the back-focal plane of a lens with $f_5$=7.5 cm. A second sequence of quarter-wave plate linear polarizer could be used in order to filter and select the desired circular-polarization state.

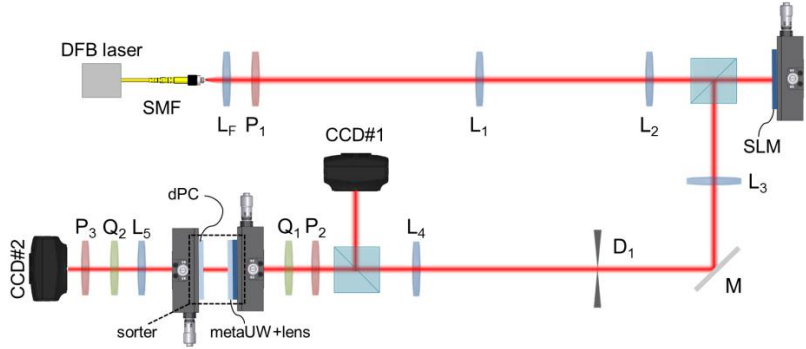

Fig. 7. Scheme of the experimental setup used for the optical characterization of the fabricated sorter. The DFB laser output (λ = 1310 nm) is collimated at the end of the single mode fiber by means of an aspheric lens ($L_F$) with a focal length $f_F$ = 7.5 mm, linearly polarized ($P_1$) and expanded ($f_1$ = 3.5 cm, $f_2$ = 10.0 cm). The SLM first order is filtered ($D_1$) and resized ($f_3$ = 20.0 cm, $f_4$ = 12.5 cm) before illuminating the sorter. A beam splitter (BS) is used both to check the input beam and collect the sorter output at the back focal plane of a fifth Fourier lens ($f_5$ = 7.5 cm). The sorter is composed of a focusing un-wrapper, realized in the form of a silicon metasurface (metaUW) coupled with a multi-level Fresnel lens (lens), and a multi-level phase-corrector (dPC). The two elements are mounted on two distinct 6-axis kinematic mounts. A sequence of linear polarizers and quarter-wave plates is placed before ($P_2$, $Q_1$) and after ($Q_2$, $P_3$) the sorter, in reverse order, in order to generate and filter the desired circular polarization states.

In Fig. 8, the far-field intensity pattern is reported for different input Laguerre-Gaussian modes and their superposition for a given circular polarization state. The optical characterization show that the sorter can discriminate both the spin and the orbital angular momentum values of light beams. The orders attributed to left-handed and right-handed circularly polarized light are clearly vertically separated while simultaneously exhibiting horizontal separation, which provides the expected discrimination between the conventional OAM orders attributed to the confocal phase elements.

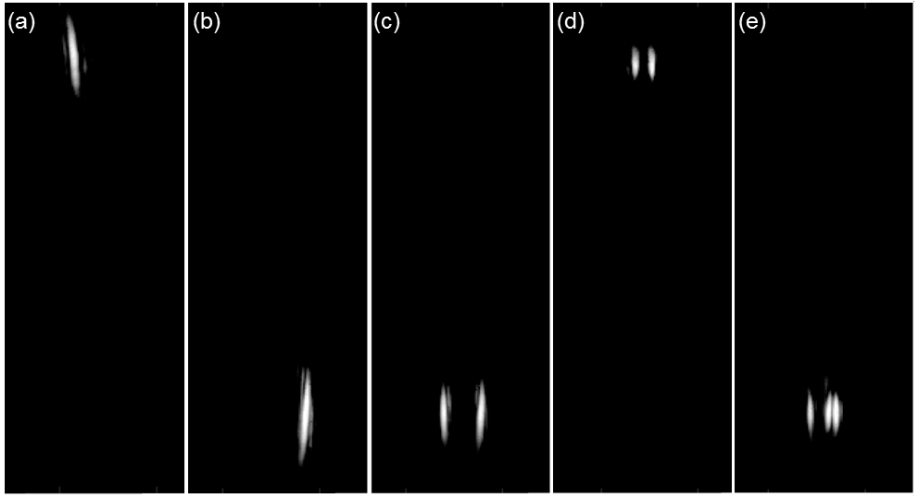

Fig. 8. Far-field intensity patterns for different input Laguerre-Gaussian beams and their superpositions: right-handed circular polarization and $\ell=-7$ (a), left-handed circular polarization and $\ell=-7$ (b), left-handed circular polarization and $\ell=-5$ & $+5$ (c), right-handed circular polarization and $\ell=-2$ & $+2$ (d), left-handed circular polarization and $\ell=-3, -1$ & $+4$ (e).

As shown in Fig. 9, it is worth noting that the transverse positions of the right-handed OAM modes are inverted with respect to those of the left-handed ones, thus verifying that these two polarizations do indeed acquire opposite phases as they propagate through the sorter. The far-field position as a function of the carried OAM is linear as expected from the theory.

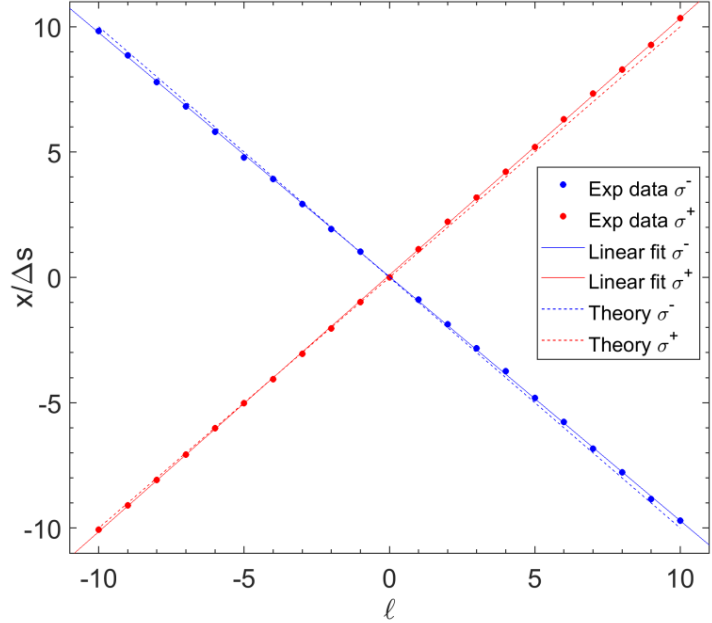

Fig. 9. Experimental positions in far-field and linear fit of the output spots as a function of the input orbital angular momentum $\ell$ for left-handed (blue) and right-handed (red) circular polarization states. The coordinate has been normalized by $\Delta s=\lambda f_5/(2\pi a)$, corresponding to the shift due to a unitary OAM increase. In our case $\Delta s=$ 39.1 μm. With this scaling, the theoretical slopes are equal +1 and -1, for right-handed and left-handed circular polarizations respectively. Experimental slopes: $+(1.02 \pm 0.01)$, $-(0.98 \pm 0.02)$.

The far-field area was split into rectangular regions centered on each elongated spot with a lateral size corresponding to the minimum separation between any two adjacent channels. By measuring the total intensity in each of these regions, we can determine the relative fraction of a specific OAM state in the input beam collected at the corresponding position in far-field or spread over the adjacent zones. In Fig. 10 the optical response is shown under illumination with pure OAM beams, generated by subsequently loading on the SLM vortex patterns with $\ell$ spanning in the range from $\ell = -10$ to $\ell = +10$ with a step $\Delta\ell = 2$ (Fig. 10).

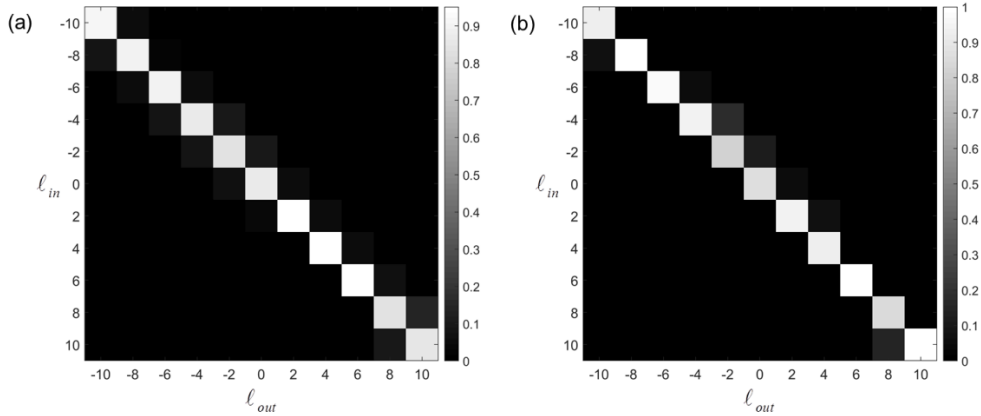

Fig. 10. Efficiency maps for input OAM beams with $\ell = -10$ to $\ell = +10$ and a step $\Delta\ell = 2$, for left-handed (a) and right-handed (b) circular polarizations.

## 5. Conclusions

In this work, we designed and fabricated, for the first time, a Pancharatnam-Berry optical element in silicon performing *log-pol* optical transformation in the telecom infrared at 1310 nm. By patterning the silicon substrate with a sub-wavelength digital grating, an inhomogeneous form-birefringence was induced, with the local fast-axis orientation equal to half the local geometric phase that should be transferred into the input beam. Due to the high gradient of the phase pattern, the sub-wavelength grating was properly engineered in order to vary continuously and produce the desired geometric phase with the highest resolution. The fabricated optical element was exploited for the demultiplexing of circularly-polarized OAM beams into a hybrid compact sorter assembled using silicon binary metasurface and multi-level diffractive optics. A Fresnel lens was coupled with the metasurface un-wrapper, in order to focus the transformed OAM-beams on the phase-corrector, independently of the input polarization states. Then, since polarization-demultiplexing is performed in the un-wrapper stage, the two polarizations impinge on the phase-corrector onto distinct zones and are processed separately by a second element fabricated in the form of multi-level diffractive optics. The optical characterization confirms the expected capability to sort polarization and OAM at the same time, providing a miniaturized total angular momentum demultiplexer that could be useful for applications in which the two degrees of freedom are exploited together, both for free-space and optical fiber propagation. For instance, OAM-modes propagating in optical fibers have been demonstrated to be inherently circularly-polarized [41], therefore the transmission and detection of guided OAM-modes cannot be exempted from polarization division demultiplexing. In addition, it could provide a promising tool for high-dimensional QKD protocols where vector-mode bases are exploited [42].

The presented method for the conversion of a phase pattern into a continuously-variant sub-wavelength grating metasurface can be extended to the design of any optical element requiring high-resolution in the phase definition. As shown is this work, if the grating period is expected to increase beyond the metasurface regime, it is enough to pixelate the whole pattern in order to impose a spatial cut-off to the grating structure. At this point, the method is limited to the control of the geometric phase. However, by introducing further degrees of freedom, also the dynamic phase could be tailored in order to either embed the focusing term in the metasurface pattern or correct chromatic aberrations [28].

**Acknowledgments**

The authors gratefully thank Dr. Giuseppe Parisi and Ing. Mauro Zontini for the interesting discussions during this work.

**Funding**

This work was supported by project New Optical Horizon region Lombardia, by project nanoMAX-nanoBRAIN of CNR, and finally by project Vortex 2 from CEPOLISPE.

**References**

1. P.J. Winzer, D.T. Neilson, and A.R. Chraplyvy, "Fiber-optic transmission and networking: the previous 20 and the next 20 years," *Opt. Express* **26**(18), 24190-24239 (2018).
2. E. Agrell, M. Karlsson, A.R. Chraplyvy, D.J. Richardson, P.M. Krummrich, P. Winzer, K. Roberts, J.K. Fisher, S.J. Savory, B.J. Eggleton, M. Secondini, F.R. Kschischang, A. Lord, J. Prat, I. Tomkos, J.E. Bowers, S. Srinivasan, M. Brandt-Pearce, N. Gisin, "Roadmap of optical communications," *J. Optics* **18**, 063002-1-40 (2016).
3. M. J. Padgett, "Orbital angular momentum 25 years on*," Opt. Express* **25**(10), 11265-11274 (2017).
4. S. Yu, "Potential and challenges of using orbital angular momentum communications in optical interconnects," *Opt. Express* **23**, 3075-3087 (2015).
5. J. Wang, "Twisted optical communications using orbital angular momentum," *China Phys. Mech. Astron*. **62**, 34201 (2019).
6. G.C.G. Berkhout, M.P.J. Lavery, J. Courtial, M.W. Beijersbergen, and M.J. Padgett, "Efficient sorting of orbital angular momentum states of light," *Phys. Rev. Lett*. **105**, 153601-1-4 (2010).
7. M.P.J. Lavery, D.J. Robertson, G.C.G. Berkhout, G.D. Love, M.J. Padgett, and J. Courtial, "Refractive elements for the measurements of the orbital angular momentum of a single photon," *Opt. Express* **20**, 2110-2115 (2012).
8. M. Mirhosseini, M. Malik, Z. Shi, and R.W. Boyd, "Efficient separation of the orbital angular momentum eigenstates of light," *Nat. Commun*. **4**, 2781 (2013).
9. C. Wan, J. Chen, and Q. Zhan, "Compact and high-resolution optical angular momentum sorter," *APL Photonics* **2**, 031302-1-6 (2017).
10. G. Ruffato, M. Massari, G. Parisi, and F. Romanato, "Test of mode division multiplexing and demultiplexing in free-space with diffractive transformation optics," *Opt. Express* **25**(7), 7859–7868 (2017).
11. G. Ruffato, M. Massari, and F. Romanato, "Compact sorting of optical vortices by means of diffractive transformation optics," *Opt. Lett*. **42**(3), 551–554 (2017).
12. G. Ruffato, M. Girardi, M. Massari, E. Mafakheri, B. Sephton, P. Capaldo, A. Forbes, and F. Romanato, "A compact diffractive sorter for high-resolution demultiplexing of orbital angular momentum beams," *Sci. Rep*. **8**, 10248 (2018).
13. F.S. Roux, "Geometric phase lens," *J. Opt. Soc. Am. A* **23**, 476-482 (2006).
14. A. Emoto, M. Nishi, M. Okada, S. Manabe, S. Matsui, N. Kawatsuki, and H. Ono, "Form birefringence in intrinsic birefringent media possessing a subwavelength structure," *App. Opt*. **49**, 4355-4361 (2010).
15. B. Desiatov, N. Mazurski, Y. Fainman, and U. Levy, "Polarization selective beam shaping using nanoscale dielectric metasurfaces," *Opt. Express* **23**, 22611-22618 (2015).
16. Z. Bomzon, V. Kleiner, E. Hasman, "Space-variant state manipulation with computer-generated subwavelength metal stripe gratings," *Opt. Comm*. **192**, 169-181 (2001).
17. Z. Bomzon, G. Biener, V. Kleiner, and E. Hasman, "Space-variant Pancharatnam–Berry phase optical elements
18. with computer-generated subwavelength gratings*," Opt. Lett*. **27**(13), 1141-1143 (2002).
19. A. Niv, G. Biener, V. Kleiner, and E. Hasman, "Propagation-invariant vectorial Bessel beams obtained by use of quantized Pancharatnam-Berry phase optical elements," *Opt. Lett*. **29**, 238-240 (2004).
20. E. Hasman, Z. Bomzon, A. Niv, G. Biener, and V. Kleiner, "Polarization beam-splitters and optical switches based on space-variant computer-generated subwavelength quasi-periodic structures," *Opt. Commun*. **209**, 45–54 (2002).
21. U. Levy, H.-C. Kim, C.-H. Tsai, and Y. Fainman, "Near-infrared demonstration of computer-generated holograms implemented by using subwavelength gratings with space-variant orientation," *Opt. Lett*. **30**, 2089-2091 (2005).

22. G. F. Walsh, "Pancharatnam-Berry optical element sorter of full angular momentum eigenstate," *Opt. Express* **24**(6), 6689-6704 (2016).
23. H. Larocque, J. Gagnon-Bischoff, D. Mortimer, Y. Zhang, F. bouchard, J. Upham, V. Grillo, R. W. Boyd, and E. Karimi, "Generalized optical angular momentum sorter and its application to high-dimensional quantum cryptography," *Opt. Express* **25**(17), 19832-19843 (2017).
24. G. F. Walsh, L. De Sio, D. E. Roberts, N. Tabiryan, F. J. Aranda, and B. R. Kimball, "Parallel sorting of orbital and spin angular momenta of light in a record large number of channels," *Opt. Lett.* **43**(10), 2256-2259 (2018).
25. S. Zheng. Y. Li, Q. Lin, X. Zeng, G. Zheng, Y. Cai, Z. Chen, S. Xu, and D. Fan, "Experimental realization to efficiently sort vector beams by polarization topological charge via Pancharatnam–Berry phase modulation," *Photon. Research* **6**(5), 385-389 (2018).
26. S. M. Choudhury, D. Wang, K. Chaudhuri, C. DeVault, A.V. Kildishev, A. Boltasseva, and V.M. Shalaev, "Material platforms for optical metasurfaces," *Nanophotonics* **7**(6), 959-987 (2018).
27. P. Genevet, F. Capasso, F. Aieta, M. Khorasaninejad, and R. Devlin, "Recent advances in planar optics: from plasmonic to dielectric metasurfaces," *Optica* **4**(1), 139-152 (2017).
28. M. Khorasaninejad, and F. Capasso, "Metalenses: Versatile multifunctional photonic components," *Science* **358**, 6367 (2017).
29. P. Capaldo, A. Mezzadrelli, A. Pozzato, G. Ruffato, M. Massari, and F. Romanato, "Nano-fabrication and characterization of silicon meta-surfaces provided with Pancharatnam-Berry effect," under publication in Optical Materials Express.
30. G. Ruffato, P. Capaldo, M. Massari, A. Messadrelli, and F. Romanato, "Pancharatnam–Berry Optical Elements for Spin and Orbital Angular Momentum Division Demultiplexing," *Photonics* **5**(4), 46 (2018).
31. Y. G. Soskind, *Field Guide to Diffractive Optics* (SPIE Press: Bellingham, Washington, USA, 2011).
32. M. G. Moharam, D.A. Pommet, E.B. Grann, T.K. Gaylord, "Stable implementation of the rigorous coupled-wave analysis for surface-relief gratings: enhanced transmittance matrix approach," *J. Opt. Soc. Am. A* **12**, 1077-1086 (1995).
33. P. Lalanne, "Improved formulation of the coupled-wave method for two-dimensional gratings," *J. Opt. Soc. Am. A* **14**, 1592-1598 (1997).
34. H. Kikuta, Y. Ohira, H. Kubo, and K. Iwata, "Effective medium theory of two-dimensional subwavelength gratings in the non-quasi-static limit," *J. Opt. Soc. Am. A* **15**(6), 1577 (1998).
35. M. Massari, G. Ruffato, M. Gintoli, F. Ricci, and F. Romanato, "Fabrication and characterization of high-quality spiral phase plates for optical applications," *Appl. Opt*. **54** (13), 4077-4083 (2015).
36. M. Beck, M. Graczyk, I. Maximov, E. L. Sarwe, T. G. I. Ling, M. Keil, and L. Montelius, "Improving stamps for 10 nm level wafer scale nanoimprint lithography," *Microelectron. Eng*. **61–62**, 441–448 (2002).
37. A. Pozzato, G. Grenci, G. Birarda, and M. Tormen, "Evaluation of a novolak based positive tone photoresist as NanoImprint Lithography resist," *Microelectron. Eng*. **88**(8), 2096–2099 (2011).
38. V. Depalma, and N. Tillman, "Friction and Wear of Self -Assembled Trichlorosilane Monolayer Films on Silicon," *Langmuir* **5**, 868–872 (1989).
39. C. Haensch, S. Hoeppener, and U. S. Schubert, "Chemical modification of self-assembled silane-based monolayers by surface reactions," *Chem. Soc. Rev*. **39**(6), 2323 (2010).
40. C. Rosales-Guzmán, A. Forbes, *How to shape light with spatial light modulators* (SPIE Press: Bellingham, Washington, USA, 2017).
41. S. Ramachandran, and P. Kristensen, "Optical vortices in fiber," *Nanophotonics* **2** (5-6), 455–474 (2013).
42. B. Ndagano, I. Nape, B. Perez-Garcia, S. Scholes, R. I. Hernandez-Aranda, T. Konrad, M.P.J. Lavery, and A. Forbes, "A deterministic detector for vector vortex states," *Sci. Rep*. **7**, 13882 (2017).